\documentclass[12pt,preprint]{aastex}

\def\hal{H$\alpha$}
\def\be{\begin{equation}}
\def\ee{\end{equation}}

\def\m{~$\mu$m}
\def\CI   {[\ion{C}{1}]}
\def\CII  {[\ion{C}{2}]}
\def\HI   {\ion{H}{1}}
\def\HII  {\ion{H}{2}}

\def\NII  {[\ion{N}{2}]}
\def\NeII {[\ion{Ne}{2}]}
\def\NeIII{[\ion{Ne}{3}]}
\def\NeV  {[\ion{Ne}{5}]}
\def\OI   {[\ion{O}{1}]}
\def\OII  {[\ion{O}{2}]}
\def\OIII {[\ion{O}{3}]}
\def\OIV  {[\ion{O}{4}]}
\def\SII  {[\ion{S}{2}]}
\def\SIII {[\ion{S}{3}]}
\def\SIV  {[\ion{S}{4}]}
\def\SiII {[\ion{Si}{2}]}
\def\ISO{{\it ISO}}

\def\Spitzer{{\it Spitzer}}
\def\2MASS{{\it 2MASS}}

\begin {document}

\title{Mid-Infrared Spectral Diagnostics of Nuclear and Extra-Nuclear Regions in Nearby Galaxies}

\author{D.A.~Dale\altaffilmark{1}, J.D.T.~Smith\altaffilmark{2}, L.~Armus\altaffilmark{3}, B.A.~Buckalew\altaffilmark{3}, G.~Helou\altaffilmark{3}, R.C.~Kennicutt\altaffilmark{4,2}, J.~Moustakas\altaffilmark{2}, H.~Roussel\altaffilmark{5,3}, K.~Sheth\altaffilmark{3}, G.J.~Bendo\altaffilmark{6,2}, D.~Calzetti\altaffilmark{7}, B.T.~Draine\altaffilmark{8},  C.W.~Engelbracht\altaffilmark{2}, K.D.~Gordon\altaffilmark{2}, D.J.~Hollenbach\altaffilmark{9}, T.H.~Jarrett\altaffilmark{3}, L.J.~Kewley\altaffilmark{10}, C.~Leitherer\altaffilmark{7}, A.~Li\altaffilmark{11}, S.~Malhotra\altaffilmark{12}, E.J.~Murphy\altaffilmark{13}, F.~Walter\altaffilmark{5}}
\altaffiltext{1}{\scriptsize Department of Physics and Astronomy, University of Wyoming, Laramie, WY 82071; ddale@uwyo.edu}
\altaffiltext{2}{\scriptsize Steward Observatory, University of Arizona, 933 North Cherry Avenue, Tucson, AZ 85721}
\altaffiltext{3}{\scriptsize California Institute of Technology, MC 314-6, Pasadena, CA 91101}
\altaffiltext{4}{\scriptsize Institute of Astronomy, University of Cambridge, Madingley Road, Cambridge CB3 0HA, United Kingdom}
\altaffiltext{5}{\scriptsize Max Planck Institut f\"{u}r Astronomie, K\"{o}nigstuhl 17, 69117 Heidelberg, Germany}
\altaffiltext{6}{\scriptsize Astrophysics Group, Imperial College, Blackett Laboratory, Prince Consort Road, London SW7 2AZ United Kingdom}
\altaffiltext{7}{\scriptsize Space Telescope Science Institute, 3700 San Martin Drive, Baltimore, MD 21218}
\altaffiltext{8}{\scriptsize Princeton University Observatory, Peyton Hall, Princeton, NJ 08544}
\altaffiltext{9}{\scriptsize NASA/Ames Research Center, MS 245-6, Moffett Field, CA 94035}
\altaffiltext{10}{\scriptsize Institute for Astronomy, 2680 Woodlawn Drive, Honolulu, HI 96822}
\altaffiltext{11}{\scriptsize Department of Physics and Astronomy, University of Missouri, Columbia, MO 65211}
\altaffiltext{12}{\scriptsize Department of Physics and Astronomy, Arizona State University, Tempe, AZ 85287}
\altaffiltext{13}{\scriptsize Department of Astronomy, Yale University, New Haven, CT 06520}

\begin {abstract}
Mid-infrared diagnostics are presented for a large portion of the Spitzer Infrared Nearby Galaxies Survey (SINGS) sample plus archival data from the {\it Infrared Space Observatory} and the {\it Spitzer Space Telescope}.  The SINGS dataset includes low- and high-resolution spectral maps and broadband imaging in the infrared for over 160 nuclear and extranuclear regions within 75 nearby galaxies spanning a wide range of morphologies, metallicities, luminosities, and star formation rates.  Our main result is that these mid-infrared diagnostics effectively constrain a target's dominant power source.  The combination of a high ionization line index and PAH strength serves as an efficient discriminant between AGN and star-forming nuclei, confirming progress made with {\it ISO} spectroscopy on starbursting and ultraluminous infrared galaxies.  The sensitivity of {\it Spitzer} allows us to probe fainter nuclear and star-forming regions within galaxy disks.  We find that both star-forming nuclei and extranuclear regions stand apart from nuclei that are powered by Seyfert or LINER activity.  In fact, we identify areas within four diagnostic diagrams containing $>$90\% Seyfert/LINER nuclei or $>$90\% \HII\ regions/\HII\ nuclei.  We also find that, compared to starbursting nuclei, extranuclear regions typically separate even further from AGN, especially for low-metallicity extranuclear environments.  In addition, instead of the traditional mid-infrared approach to differentiating between AGN and star-forming sources that utilizes relatively weak high-ionization lines, we show that strong low-ionization cooling lines of X-ray dominated regions like \SiII~34.82\m\ can alternatively be used as excellent discrimants.  Finally, the typical target in this sample shows relatively modest interstellar electron density ($\sim 400~{\rm cm}^{-3}$) and obscuration ($A_V \sim 1.0$~mag for a foreground screen), consistent with a lack of dense clumps of highly obscured gas and dust residing in the emitting regions.
\end {abstract}
 
 
\section {Introduction}
\label{sec:intro}

The goal of this study is to explore whether mid-infrared diagnostics developed for luminous/ultraluminous infrared galaxies and bright Galactic \HII\ regions can be improved upon and extended to the nuclear and extra-nuclear regions within normal and infrared-faint galaxies.  A traditional method for characterizing a galaxy's nuclear power source uses ratios of optical emission lines such as \OII~3727\AA, H$\beta$~4861\AA, \OIII~5007\AA, \OI~6300\AA, H$\alpha$~6563\AA, \NII~6584\AA\, and \SII~6716,6731\AA\ (e.g., Baldwin, Phillips, \& Terlevich 1981; Veilleux \& Osterbrock 1987; Ho, Filippenko, \& Sargent 1997a; Kewley et al. 2001; Kauffmann et al. 2003).  A plot of \OIII/H$\beta$ versus \NII/H$\alpha$, for example, will typically separate Seyferts, LINERs, and starburst nuclei.  Since nuclei are often heavily enshrouded by dust, especially in luminous and ultraluminous infrared galaxies
(LIRGs and ULIRGs), an important limitation to galaxy optical diagnostics is the effect of extinction.  In anticipation of the data stream from space-based infrared platforms, early theoretical work with photoionization models showed that infrared ionic fine structure line ratios could profitably enable astronomers to approach galaxy classification from a new perspective (e.g., Voit 1992; Spinoglio \& Malkan 1992).  The advent of sensitive infrared line data from the {\it Infrared Space Observatory} was an important first step to peering more deeply into buried nuclear sources (Genzel et al. 1998; Laurent et al. 2000; Sturm et al. 2002; Peeters, Spoon, \& Tielens 2004).  Genzel and collaborators were the first to show that ionization-sensitive indices based on mid-infrared line ratios correlate with the strength of polycyclic aromatic hydrocarbon (PAH) emission features.  AGN in particular show weak PAH and large ratios of high-to-low ionization line emission.  Interestingly, while Genzel et al. found that 70-80\% of their ULIRG sample is mainly powered by starburst activity, the percentage appears to drop for higher luminosity ULIRGs (e.g., Veilleux, Sanders, \& Kim 1999, but see also Farrah et al. 2003).  Taniguchi et al. (1999) and Lutz, Veilleux, \& Genzel (1999) suggest that optical and infrared classifications agree if the nuclei of LINER-like ULIRGs are in fact dominated by shocks driven by powerful supernova winds.  Mid-infrared lines observed by \ISO\ were also used to probe the physical characteristics and evolution of purely starbursting nuclei (Thornley et al. 2000; Verma et al. 2003) and Galactic \HII\ regions (Vermeij \& van der Hulst 2002; Giveon et al. 2002).  An important result stemming from these efforts is that stellar aging effects appear to result in \HII\ regions generally having higher excitations than starbursting nuclei.

The unprecedented sensitivity and angular resolution afforded by the {\it Spitzer Space Telescope} allow an even more detailed view into the nature of galaxy nuclei (e.g., Armus et al. 2004; Smith et al. 2004).  SINGS takes full advantage of {\it Spitzer}'s capabilities by executing a comprehensive, multi-wavelength survey of 75 nearby galaxies spanning a wide range of morphologies, metallicities, luminosities, and star formation activity levels (Kennicutt et al. 2003).  The sensitivity of {\it Spitzer} coupled with the proximity of the SINGS sample allows dwarf galaxy systems fainter than $L_{\rm FIR} \sim 10^7~L_\odot$ to be spectroscopically probed in the infrared for the first time.  In addition, prior to {\it Spitzer} the only individual extragalactic \HII\ regions that were detectable with infrared spectroscopy resided in the Local Group (e.g., Giveon et al. 2002; Vermeij et al. 2002).  In contrast, SINGS provides infrared spectroscopic data for nearly 100 extragalactic \HII\ regions, residing in systems as near as Local Group members to galaxies as far as $\sim$25~Mpc.  The SINGS dataset thus samples a wider range of environments than previously observed with infrared spectroscopy.  This diversity in the SINGS sample provides a huge range for exploring physical parameters with mid-infrared spectral diagnostics.  The high ionization lines historically used in such diagnostics, such as \OIV~25.89\m\ and \NeV~14.32\m, are relatively weak and can be difficult to detect in lower luminosity systems.  Fortunately, high ionization lines are not the only route to determining whether a galaxy harbors a strong AGN.  Similar to how the \OI~6300\AA\ and \OI~63\m\ lines are AGN diagnostics (e.g., Dale et al. 2004a), we show below the utility of using the comparatively bright \SiII~34.82\m\ mid-infrared line as another effective tool for deciphering a galaxy's power source.  

\section {The Sample}

\subsection {Galaxy Nuclei}

The sample of nuclear targets analyzed in this study derive from the SINGS Third Data Release.  These 50 nuclei come from a wide range of environments and galaxies: low-metallicity dwarfs; quiescent ellipticals; dusty grand design spirals; Seyferts, LINERs, and starbursting nuclei of normal galaxies; and systems within the Local and M~81 groups (see Kennicutt et al. 2003).  

\subsection {Extra-Nuclear Regions}

The 26 extranuclear sources studied in this work also come from the SINGS Third Data Release.  These targets stem from the original set of 39 optically-selected sources listed by Kennicutt et al. (2003).  The optically-selected OB/\HII\ regions cover a large range of metallicity (0.1-3~$Z_\odot$), extinction-corrected ionizing luminosity (10$^{49}-10^{52}$~photons~s$^{-1}$), extinction ($A_V \lesssim 4$~mag), radiation field intensity (ionization parameter $\log U=-2$ to $-4$; Habing 1968), ionizing stellar temperature ($T_{\rm eff}=35-55$~kK), and local H$_2$/\HI\ ratio as inferred from CO ($<$0.1 to $>$10).  Additional extranuclear targets for the SINGS project  have since been identified, based on their infrared properties, but their observations necessarily came later than the observations of the optically-selected targets.  Those ``Second Look'' data will be the focus of a future paper.

\section {The Data}

The full SINGS observing program and the data processing are described by Kennicutt et al. (2003), Smith et al. (2004), and Dale et al. (2005).  Here we briefly summarize the spectral observations and data processing relevant to this paper.  

\subsection {{\it Spitzer} Infrared Spectroscopy Observations and Data Processing}
\label{sec:IRobservations}

\begin{figure}
 \plotone{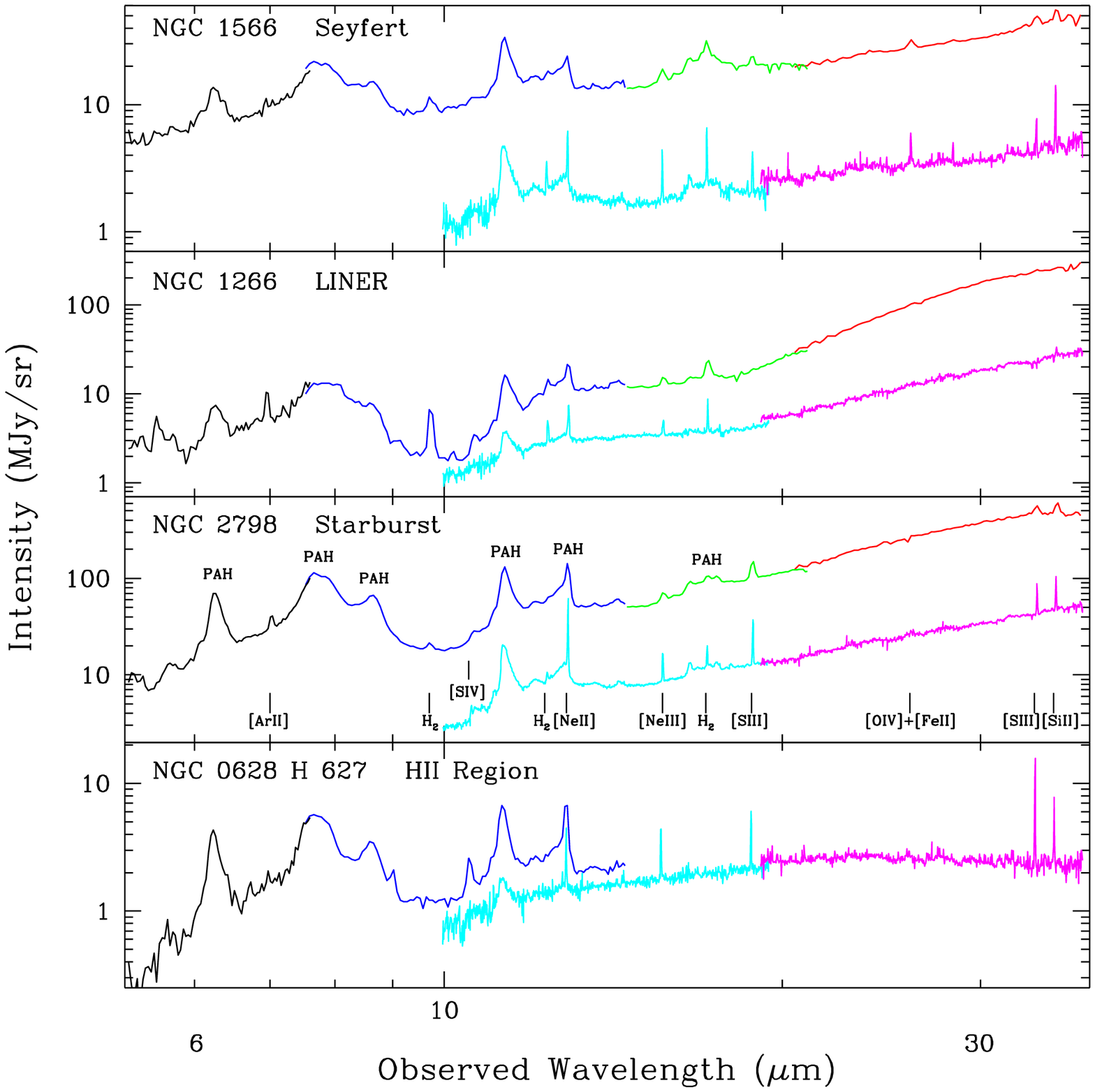}
 \caption{Examples of low- and high-resolution spectra for four different types of environments found within the SINGS sample.  The IRS modules span wavelengths $\sim$5-7\m\ (Short-Low2), $\sim$7-14\m\ (Short-Low1), $\sim$14-21\m\ (Long-Low2), $\sim$21-40\m\ (Long-Low1), $\sim$10-19\m\ (Short-High), and $\sim$19-37\m\ (Long-High).  Each spectral segment has been extracted from a $\sim$23\arcsec$\times$15\arcsec\ region.  The high-resolution data are scaled downwards by 1.0~dex for clarity.  The variable offset between the low- and high-resolution data is largely due to the lack of sky subtraction for the high-resolution data.}
 \label{fig:spectra}
\end{figure}

High-resolution spectroscopy ($R\sim600$) was obtained in the Short-High (10-19\m) and Long-High (19-37\m) modules, and low-resolution spectroscopy ($R\sim50-100$) was obtained in the Short-Low (5-14\m) and Long-Low (14-38\m) modules (Houck et al. 2004a).  Figure~\ref{fig:spectra} shows example spectra for a variety of sources (Long-Low data are not used elsewhere in this work).  Nuclei were generally mapped with a 3$\times$5 grid (Short-High and Long-High) and a 1$\times$18 grid (Short-Low), utilizing half-slit width and half-slit length steps.  Extra-nuclear targets were observed with a similar scheme but with a 1$\times$9 Short-Low grid.  For a subset of nine sources with extended circumnuclear star formation we obtained slightly larger (6$\times$10) Short-High nuclear maps.  Owing to the different angular sizes subtended by the instruments, the resulting maps are approximately 57\arcsec$\times$31\arcsec\ and 57\arcsec$\times$18\arcsec\ in Short-Low (nuclear  and extra-nuclear, respectively), 45\arcsec$\times$33\arcsec\ in Long-High, and 23\arcsec$\times$15\arcsec\ in Short-High (the 10 expanded Short-High nuclear maps are 40\arcsec$\times$28\arcsec).  All integrations are 60~s per pointing, except the Short-Low nuclear maps are 14~s per pointing.  The effective integrations are longer since each location was covered 2-4 times.

The individual data files for a given spectral map were assembled into spectral cubes using the software CUBISM (Kennicutt et al. 2003; Smith et al. 2006, in preparation).  The cube input data were pre-processed using version S12.0 of the {\it Spitzer Science Center} pipeline.  Various post-processing steps within CUBISM are described by Smith et al. (2004).  Short-Low sky subtraction was enabled via the extended off-source wings of the Short-Low module (and occasionally using spatially- and temporally-proximate data from the archive when our off-source wings do not extend to the sky).  Several cross-checks on the flux calibration were made between Short-Low, Short-High, Long-Low, Long-High, MIPS, and IRAC data.  The absolute flux calibration uncertainty for the spectral data is estimated to be 25\%; the uncertainy in line flux ratios is $\sim$10\%.

Though the Short-High, Long-High, and Short-Low cubes all span different solid angles, the same matched extraction apertures were used for all cubes: one-dimensional spectra were extracted from the three-dimensional data using $\sim$23\arcsec$\times$15\arcsec\ apertures.  Furthermore, the extraction apertures are centered on the optically-derived coordinates listed by Kennicutt et al. (2003); 
the optical coordinates generally coincide with the infrared emission peak.
Emission line and PAH feature fluxes and equivalent widths are derived from continuum-subtracted Gaussian fits to the lines and first- or second-order polynomial fits to the continua.


\subsection {Archival Spectroscopy}
\label{sec:ISO}

\ISO-SWS and \Spitzer\ IRS line fluxes are drawn from the literature for a wide variety of sources including Galactic, Magellanic Cloud, and Local Group \HII\ regions (Vermeij et al. 2002; Giveon et al. 2002; Peeters et al. 2002), and starburst and active galaxies (Genzel et al. 1998; Sturm et al. 2002; Verma et al. 2003; Armus et al. 2004; Peeters, Spoon, \& Tielens 2004; Weedman et al. 2005; Haas et al. 2005).  Equivalent widths of the 6.2\m\ PAH feature were extracted from archival \ISO-PHOT and \Spitzer\ IRS data, when available.  Note that, for a given source, only data from \Spitzer\ or only data from \ISO\ are used; cross-observatory data are not used in this analysis. The fields-of-view of the \ISO-PHOT and \ISO-SWS apertures (respectively 24\arcsec$\times$24\arcsec\ and 14\arcsec$\times$20\arcsec\ to 20\arcsec$\times$30\arcsec) provide a reasonable match to the $\sim$23\arcsec$\times$15\arcsec\ extraction apertures described in \S~\ref{sec:IRobservations}.

\subsection {{\it Spitzer} Broadband Imaging: 24\m}

The SINGS project did not obtain nearby sky observations in HighRes mode, so if the underlying hot dust continuum emission is detected, the foreground/background continuum is not subtracted.  SINGS IRS HighRes spectroscopy is designed to measure line {\it fluxes} and thus line equivalent width measures via HighRes are biased by the un-subtracted sky emission.  On the other hand, the SINGS program includes extensive MIPS (Rieke et al. 2004) 24\m\ broadband imaging of all galaxies; the 24\m\ data can be used to normalize line fluxes.  Sky-subtracted MIPS 24\m\ fluxes are extracted from aperture cutouts matched to the Short-High field-of-view (\S~\ref{sec:IRobservations}).  See Kennicutt et al. (2003) and Dale et al. (2005) for further details of the SINGS broadband imaging.

\subsection {Optical Spectroscopy Observations and Data Processing}

Optical spectrophotometry has been obtained for the SINGS project at the Steward Observatory Bok 2.3~m and the CTIO 1.5~m telescopes (Moustakas \& Kennicutt 2006).  A suite of spectral drift scans, centered on the nuclei, were taken to spatially map the various regions covered with the IRS spectroscopy program (see Kennicutt et al. 2003 for more details).  The optical spectroscopy has spectral resolution of $\sim$8\AA\ and covers 3600-7000\AA\ so that the primary nebular emission lines can be studied (e.g., \OII~3727\AA, H$\beta$~4861\AA, \OIII~5007\AA, H$\alpha$~6563\AA, \NII~6584\AA).  The SINGS optical spectroscopy and emission line 
measurements (fluxes, equivalent widths, metallicities, etc.) will be presented by Moustakas et al. (2006, in preparation).  

This work utilizes optical spectral drift scans integrated over the central 
20\arcsec$\times$20\arcsec\ regions, 
approximately matching the $\sim$23\arcsec$\times$15\arcsec\ circumnuclear regions 
over which the infrared line and PAH feature fluxes are extracted (see \S~\ref{sec:IRobservations}).  

\section {Measured Quantities}

\subsection {High-Resolution Infrared Spectroscopy: Emission Lines}

Several forbidden lines are quite prominent in many of the SINGS nuclear and extra-nuclear high resolution spectra (\NeII12.81\m, \NeIII15.56\m, \SIII18.71\m, \SIII33.48\m, and \SiII34.82\m) along with a few other higher ionization
lines that are occasionally observed (\SIV10.51\m, \NeV14.32\m, and \OIV25.89\m; Figure~\ref{fig:spectra}).
All these lines except \SiII34.82\m\ are nebular lines from hydrogen gas ionized regions; \SiII34.82\m\ comes from a wider variety of regions, including both ionized gas and warm atomic gas such as photodissociation regions and X-ray dominated regions (e.g., Hollenbach \& Tielens 1999).
A series of interstellar molecular hydrogen lines are also detected in the SINGS spectra; these are explored in a separate paper (Roussel et al. 2006, in preparation).  The line fluxes utilized in this work are listed in Tables~\ref{tab:fluxes_nuc} and \ref{tab:fluxes_extranuc}.  Eight galaxies from the SINGS Third Data Release are not listed in Table~\ref{tab:fluxes_nuc} because their nuclear regions were too faint to be observed (or detected if observed).

\subsection {Low-Resolution Infrared Spectroscopy: The 6.2\m\ PAH Feature}

The strength of a PAH 
feature depends on a complex combination of several parameters of the interstellar medium, some of which are interlinked: metallicity, dust column density, the distribution of sizes and ionization states in the dust grain population, and the intensity and hardness of the interstellar radiation field (e.g., Cesarsky et al. 1996; Thuan et al. 1999; Sturm et al. 2000; Li \& Draine 2001; Draine \& Li 2001; Houck et al. 2004b; Galliano et al. 2005; Engelbracht et al. 2005; Madden et al. 2006; Wu et al. 2006).  Due to this sensitivity to the properties of the interstellar medium, PAH features in the mid-infrared have been used to characterize the physical state 
of a system.
Previous efforts have utilized the 6.2\m\ feature (Laurent et al. 2000; Peeters, Spoon, \& Tielens 2004; Weedman et al. 2005), the 7.7\m\ feature (Genzel et al. 1998), or a combination of several PAH features (e.g., Verstraete et al. 2001; Tran et al. 2001; Peeters et al. 2002; F\"orster Schreiber et al. 2004; Armus et al. 2004; Peeters et al. 2004; Madden et al. 2006).  For this work we concentrate on diagnostics that utilize the 6.2\m\ feature.  The 6.2\m\ feature is the only strong infrared signature of PAHs not blended with an emission line or absorption trough, not near the wavelength edges of an IRS low resolution module, and for which the blue and red sides of the ``continuum'' are easy to define.  The equivalent widths of the 6.2\m\ feature are listed in Tables~\ref{tab:fluxes_nuc} and \ref{tab:fluxes_extranuc}.  The equivalent width is used in our diagnostics since 
the continuum at 6.2\m\ may contain emission from hot dust heated either by star formation or an AGN, in addition to a fractional contribution from stellar emission; the 6.2\m\ equivalent width is (indirectly) sensitive to the presence of an AGN.

\section {Results}

\subsection {Optical Classifications of Nuclei}
\label{sec:optical}

\begin{figure}
 \plotone{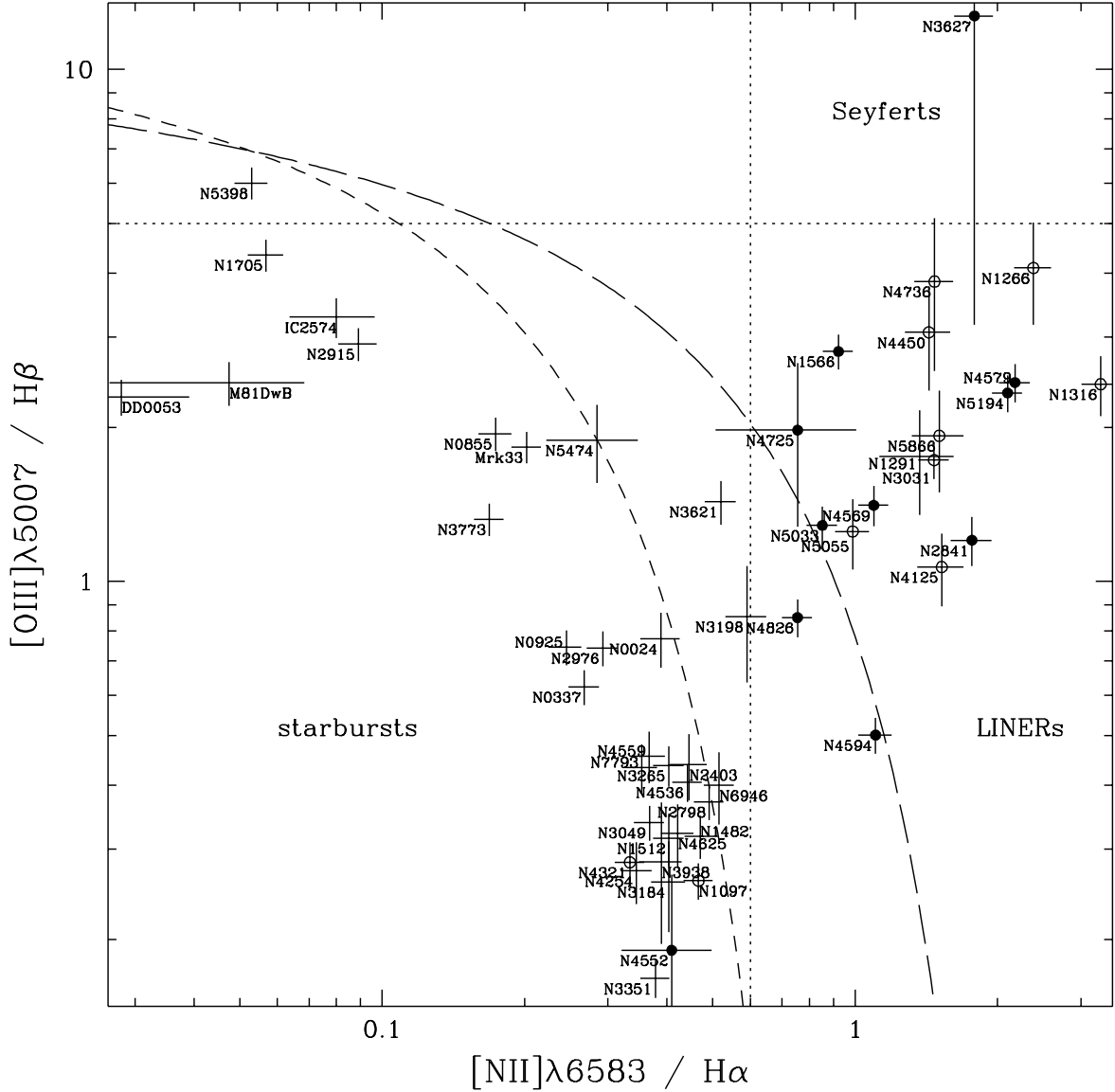}
 \caption{A traditional diagnostic diagram is displayed for the SINGS nuclei using optical data and 20\arcsec$\times$20\arcsec\ apertures (uncorrected for reddening).  Filled (open) circles mark galaxies for which the literature indicates a Seyfert (LINER) nucleus.  The dotted lines delineate typical starburst/Seyfert/LINER boundaries: \OIII 5007/H$\beta \sim5$ and \NII 6583/H$\alpha \sim0.6$ (e.g., Armus, Heckman, \& Miley 1989).  The long-dashed and short-dashed curves respectively trace the starburst/AGN boundaries of Kewley et al. (2001; theoretical) and Kauffmann et al. (2003; empirical).  Error bars represent 1$\sigma$ uncertainties.}
 \label{fig:bpt}
\end{figure}

Not all of the SINGS galaxies have an optical nuclear classification in the literature (e.g., Seyfert, LINER, starburst, etc).  Moreover, online databases such as NED\footnote{NASA/IPAC Extragalactic Database} provide a heterogeneous source for such classifications.  Thus we turn to our own optical spectroscopy.  Figure~\ref{fig:bpt} displays a traditional diagnostic diagram (e.g., Baldwin, Phillips, \& Terlevich 1981) using optical spectroscopy and 
20\arcsec$\times$20\arcsec\
apertures for SINGS galaxy nuclei.  
Note that such diagnostics were designed with ratios of lines closely spaced in wavelength to minimize the effects of extinction.
Filled (open) circles in Figure~\ref{fig:bpt} mark galaxies for which the literature indicates a Seyfert (LINER) nucleus; data points without circles in this diagram represent galaxies without a LINER or Seyfert classification in the literature.  As alluded to in \S~\ref{sec:intro}, the LINER classification can be complicated.  Similar to what is observed for Seyferts, the optical properties of LINERs are consistent with a hard power-law spectrum.  But LINER-type spectra can also be produced via winds, shocks, and cooling flows (Kauffmann et al. 2003).  To further complicate the picture, ``transition'' objects are thought to be LINER or Seyfert galaxies with substantial contributions from normal star formation (e.g., Ho, Filippenko, \& Sargent 1993; Gonz\'alez Delgado et al. 2004).

The dotted lines delineate typical starburst/AGN/LINER boundaries: \OIII 5007/H$\beta \sim5$ and \NII 6583/H$\alpha \sim0.6$ (e.g., Veilleux \& Osterbrock 1987; Armus, Heckman, \& Miley 1989).  The long-dashed curve traces the theoretical starburst/AGN boundary of Kewley et al. (2001), marking the maximum position in this diagram that can be obtained by pure photoionization.
Objects lying above this curve require an additional power source such as an AGN or shocks; objects lying below this curve  may still contain an AGN responsible for up to $\sim$30\% of the emission line flux ratios.  The short-dashed curve traces an empirical starburst/AGN boundary based on data from tens of thousands of Sloan Digital Sky Survey galaxies (Kauffmann et al. 2003).  The Kauffman et al. curve aims to define a boundary below which no galaxies contain an AGN.
Objects lying in between the Kewley et al. and the Kauffmann et al. curves are likely to be composite AGN/starburst objects but still dominated by star formation.  It is evident that the classification information available from the literature is insufficient or incomplete for a handful of SINGS nuclei; there is overall good agreement between the literature and our classifications, with a few exceptions.  
The literature
does not indicate that the nuclei of NGC~1291 
contains LINER or Seyfert activity, whereas our spectroscopic data place 
it squarely in the LINER category.
On the other hand, NED lists NGC~1097, NGC~4321, and NGC~4552
as having LINER or Seyfert nuclei, yet our optical line ratios suggest they are 
star-forming galaxies.
NGC~3198, NGC~3621, and NGC4826 appear to lie in a transitional regime between the starburst and LINER/Seyfert regions.\footnote{Similar results are found using \SII~6716,6731\AA\ and \OI~6300\AA\ data in place of \NII~6584\AA\ (e.g., Kewley et al. 2001).}

Using the H$\alpha$/H$\beta$ ratio extracted from our optical spectra and a screen model for the dust distribution within a galaxy, the SINGS nuclei show modest attenuations, $\left<A_V\right> \sim 1.0$~mag with a dispersion of 1.0~mag and a maximum of $A_V \sim4.1$~mag for NGC~1266.  No sources appear to be heavily buried in the optical so presumably none of the classifications are skewed by large amounts of dust.\footnote{The exception is NGC~1377, a deeply obscured system for which the optical data are likely probing only outer layer (foreground) emission (Roussel et al. 2003, 2006).}  This result is consistent with the lack of deeply buried objects in comparisons of SINGS \hal\ and \Spitzer\ 24\m\ data (e.g., Calzetti et al. 2005).  However, as described above, there may be a few transitional objects for which the nuclear classifications are difficult to interpret in the optical.  The main reason behind the analysis in this section is to provide classifications for sources that do not yet have them from the literature.  Our goal is not to carry out a detailed analysis of the relative merits of classifying in the optical versus 
in the infrared, primarily since the SINGS sample is not optimally suited for such a test.
In the next section we turn to exploring new and existing infrared diagnostics.

\subsection {Infrared Spectral Diagnostics of Nuclei and Extra-Nuclear Regions}
\subsubsection {Emission Line Ratios and PAH Strength}
\label{sec:lines_pah}

\begin{figure}
 \plotone{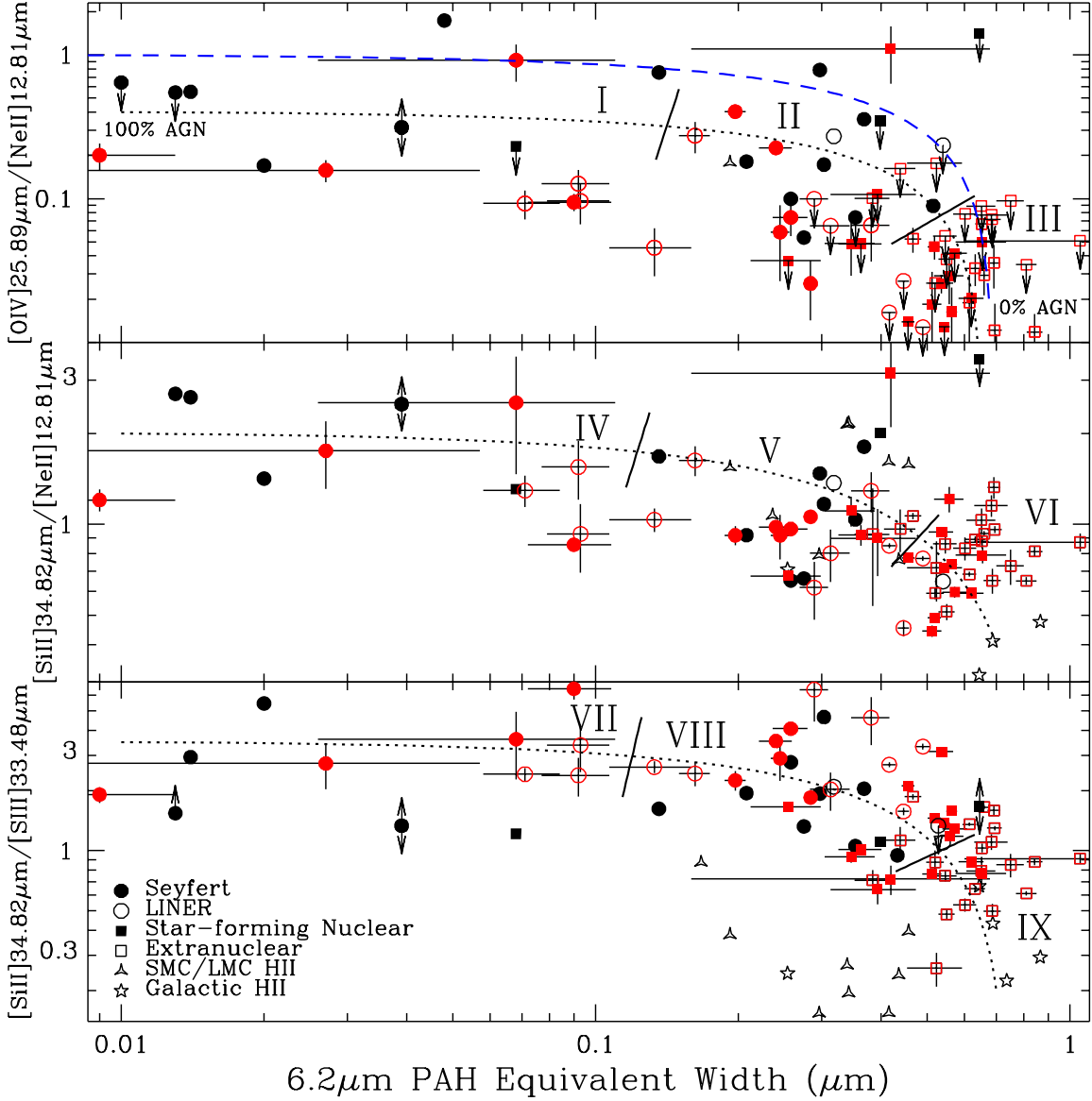}
 \caption{The ratios of mid-infrared forbidden lines 
 as a function of the 6.2\m\ PAH feature equivalent width.  SINGS data are displayed in red with 1$\sigma$ error bars based on the statistical uncertainties; archival data without error bars are indicated with black symbols and described in \S~\ref{sec:ISO}.  The dotted lines are linear mixing models of 
a ``pure'' AGN and a ``pure'' star-forming source
 (see text).  The dashed line in the top panel is a similar mixing model first presented by Genzel et al. (1998).  The solid lines and Roman numerals delineate regions distinguished by Seyferts, LINERs, star formation, etc. (see Table~\ref{tab:lines_pah} and \S~\ref{sec:lines_pah}).}
 \label{fig:lines_pah}
\end{figure}

At least 54.9~eV is required to remove an electron from doubly-ionized oxygen.  
On the other hand, ionizing neutral neon ``only'' requires 21.6~eV.  Compared to 21.6~eV photons, 54.9~eV photons are far more likely to stem from accretion-powered disks than star formation (e.g., OB stars; Smith et al. 2004), and thus the ratio of \OIV 25.89\m\ and \NeII 12.81\m\ depends on the type of source dominating the energetics of the interstellar medium.  Furthermore, studies show that PAH features are quite prominent throughout much of the interstellar medium except for regions characterized by exceptionally hard radiation fields such as those that arise in AGN and the cores of \HII\ regions (Cesarsky et al. 1996; Sturm et al. 2000).  A mid-infrared diagnostic diagram first put forth by Genzel et al. (1998), and later explored by Peeters, Spoon, \& Tielens (2004), plots the emission line ratio \OIV 25.89\m/\NeII 12.81\m\ versus the strength of a mid-infrared PAH feature.  In such plots AGN sources show enhanced \OIV 25.89\m\ emission and comparatively weak PAH feature strength.  However, those two studies focussed on AGN-dominated, ULIRG, and starburst systems.  The upper panel of Figure~\ref{fig:lines_pah} uses the 6.2\m\ PAH feature and the emission line ratio \OIV 25.89\m/\NeII 12.81\m\ in a similar mid-infrared diagnostic diagram, but one that incorporates ``normal'' (starbursting/star-forming) nuclei and \HII\ regions---the sensitivity of \Spitzer\ allows us to probe to far fainter levels than heretofore possible.\footnote{Additional high ionization line possibilities include \NeV 14.32\m\ and \SIV 10.51\m, but these are less frequently detected than \OIV 25.89\m\ in SINGS spectra.}  As expected, normal nuclei and \HII\ regions extend the previously-observed trend: lower-luminosity star-forming nuclei and \HII\ regions exhibit comparatively large 6.2\m\ equivalent widths and relatively low ratios of \OIV 25.89\m/\NeII 12.81\m, indicating strong contributions from \NeII 12.81\m\ cooling of \HII\ regions and their PAH-rich photodissociation region surroundings, and negligible emission from AGN.  

High ionization lines like \OIV 25.89\m\ are somewhat difficult to detect in many SINGS sources.  Two alternative diagnostic diagrams are also provided in Figure~\ref{fig:lines_pah}.  Both the middle and lower panels involve the more easily detectable \SiII 34.82\m\ line, which has an ionization potential of 8.15~eV.  The normalization for the \SiII 34.82\m\ line in the middle and lower panels utilizes strong mid-infrared lines with similar ionization potentials: \NeII 12.81\m\ (21.6~eV) and \SIII 33.48\m\ (23.3~eV).  An advantage to using \NeII 12.81\m\ as the normalization is that it lies in the much less noisy Short-High module of IRS.  Conversely, the \SIII 33.48\m\ line lies in the same Long-High module as the \SiII 34.82\m\ line, which can improve observing efficiency and minimize cross-module uncertainties involving calibration and aperture matching.  In addition, the short wavelength baseline between the \SiII 34.82\m\ \SIII 33.48\m\ lines minimizes the effects of extinction.

Why are similar trends seen in the three panels?  The answer may lie in the physics of X-ray dominated regions.  As pointed out by Maloney, Hollenbach, \& Tielens (1996), the \SiII 34.82\m\ line is a strong coolant of X-ray irradiated gas.  In X-ray dominated regions around AGN, the \SiII\ emission dominates that from the comparatively small \HII-like regions surrounding the hard-spectrum source.  Moreover, X-ray dominated regions can be quite large since hard X-ray photons penetrate large column densities, and the conversion of X-ray energy to infrared continuum and line emission can be very efficient.  Maloney, Hollenbach, \& Tielens (1996) predict that \SiII 34.82\m, \OI 63\m, \CII 158\m, and \CI 609\m\ are the top four cooling lines within X-ray dominated regions, with \SiII34.82\m\ having an amplitude 1-10\% that of the far-infrared luminosity for an extremely large range of physical conditions.  

An argument based on interstellar density provides another possibility for the high \SiII 34.82\m/\NeII 12.81\m\ and SiII 34.82\m/\SIII 33.48\m\ ratios in AGN.   Kaufman, Wolfire, \& Hollenbach (2006) show that the ratio \SiII(PDR)/\SiII(\HII) increases with increasing density.  In fact, for low density \HII\ regions most of the \SiII\ comes from the \HII\ region and not the surrounding photodissociation region.  Moreover, Meijerink \& Spaans (2005) show that the ratio \SiII(XDR)/\SiII(PDR) also increases with increasing density.  AGN may have their emitting gas at higher densities than typically found in starbursts and normal galaxies, leading to increased \SiII 34.82\m/\NeII 12.81\m\ and SiII 34.82\m/\SIII 33.48\m\ ratios.  In other words, the prominent \SiII 34.82\m\ line for AGN sources may be due to strong \SiII\ cooling of X-ray dominated regions or enhanced \SiII\ emission from the surrounding dense photodissociation regions.  

A third scenario for enhanced \SiII 34.82\m\ in AGN involves the extent to which silicon is depleted onto dust grains.  Heavy elements such as Si, Mg, and Fe may be returned to the gas phase by dust destruction (e.g., sputtering) in regions subject to strong shocks caused by stellar winds, starbursts, and AGN activity.  So perhaps (gas phase) silicon lines are stronger in active galaxies due to this effect.

If the strong \SiII 34.82\m\ emission is due to the cooling of X-ray dominated regions, it should be noted that relatively strong low-ionization line emission (e.g., \OI 6300\AA\ and \OI 63\m) has previously been observed emanating from the large ``partially-ionized regions'' surrounding AGN and infrared-bright galaxies (Veilleux \& Osterbrock 1987; Armus, Heckman, \& Miley 1989; Veilleux 1991; Spinoglio \& Malkan 1992; Osterbrock 1993; Dale et al. 2004a).  Hence, strong low-ionization line emission from AGN is not a new concept.  We take advantage of this concept to present new techniques for distinguishing between AGN sources and star-forming regions.  These techniques rely on an easily-detectable, prominent cooling line of a low-ionization species associated with X-ray dominated regions, the dense interstellar material illuminated by power-law radiation fields.

The regions of Figure~\ref{fig:lines_pah} where Seyferts/LINERs/starbursts mix are quite large, though the bottom panel perhaps shows a cleaner separation (less mixing) between Seyferts+LINERs and star-forming regions; only the top left and bottom right extremes allow for a clean separation between classifications.  Short solid lines roughly perpendicular to the dotted AGN/star-forming curves delineate three regions in the panels of Figure~\ref{fig:lines_pah}.  The boundaries are:
\begin{eqnarray}
{\rm Region [~I-~II]}:   \log ( [{\rm O~IV}]25.89\micron  / [{\rm Ne~II}]12.81\micron ) &=& 10~ \log ({\rm EW}[6.2\micron {\rm PAH}]) + 8.0\\
{\rm Region [II-III]}:   \log ( [{\rm O~IV}]25.89\micron  / [{\rm Ne~II}]12.81\micron ) &=& 1.9 \log ({\rm EW}[6.2\micron {\rm PAH}]) - 0.6\\
{\rm Region [IV-V]}:     \log ( [{\rm Si~II}]34.82\micron / [{\rm Ne~II}]12.81\micron ) &=& 5.0 \log ({\rm EW}[6.2\micron {\rm PAH}]) + 4.8\\
{\rm Region [V-VI]}:     \log ( [{\rm Si~II}]34.82\micron / [{\rm Ne~II}]12.81\micron ) &=& 1.7 \log ({\rm EW}[6.2\micron {\rm PAH}]) + 0.5\\
{\rm Region [VII-VIII]}: \log ( [{\rm Si~II}]34.82\micron / [{\rm S~III}]33.48\micron ) &=& 10~ \log ({\rm EW}[6.2\micron {\rm PAH}]) + 9.7\\
{\rm Region [VIII-IX]}:  \log ( [{\rm Si~II}]34.82\micron / [{\rm S~III}]33.48\micron ) &=& 1.1 \log ({\rm EW}[6.2\micron {\rm PAH}]) + 0.3
\end{eqnarray}
The population statistics for these regions, provided in Table~\ref{tab:lines_pah}, show that Regions~[I+IV+VII] and [III+VI+IX] are respectively representative (at the $>$90\% level) of Seyferts/LINERs and star-forming systems.  Regions [II+V+VIII], on the other hand, contain a mix of classifications and thus represent transition regions---either the source classifications in this region are ambiguous or the region simply contains a more heterogeneous mixture of pure types.  Seyfert nuclei could shift toward the location of star-forming nuclei due to aperture effects---though the same solid angles are used for extracting the line data, the range of distances in the sample leads to a range in physical apertures.  Conversely, some star-forming nuclei exhibit relatively large line ratios and small PAH equivalent widths in Figure~\ref{fig:lines_pah}.  Perhaps a fraction of the star-forming regions contain significant numbers of Wolf Rayet stars, leading to enhanced \OIV\ emission (e.g., Schaerer \& Stasi\`nska 1999).  And maybe a decreased PAH equivalent width results from a relatively low ratio of photodissociation region to \HII\ region contributions (Laurent et al. 2000).  

The dotted lines in Figure~\ref{fig:lines_pah} represent a variable mix of an AGN nucleus and a ``pure'' star-forming region.  The anchors for these mixing models are the following:
\begin{eqnarray}
{\rm EW}[6.2\micron {\rm PAH}]\approx0.01 \micron,  &&\;\;\; [{\rm O~IV}]25.89\micron  / [{\rm Ne~II}]12.81\micron  \approx0.4  \;\;\;\;\;\;\;\;\; 100\%~{\rm AGN}  \nonumber \\
{\rm EW}[6.2\micron {\rm PAH}]\approx0.7  \micron,~ &&\;\;\; [{\rm O~IV}]25.89\micron  / [{\rm Ne~II}]12.81\micron  \approx0.01 \;\;\;\;\;\;\; 100\%~{\rm H~II} \nonumber \\
{\rm EW}[6.2\micron {\rm PAH}]\approx0.01 \micron,  &&\;\;\; [{\rm Si~II}]34.82\micron / [{\rm Ne~II}]12.81\micron~ \approx2    \;\;\;\;\;\;\;\;\;\;\;\; 100\%~{\rm AGN}  \nonumber \\
{\rm EW}[6.2\micron {\rm PAH}]\approx0.7  \micron,~ &&\;\;\; [{\rm Si~II}]34.82\micron / [{\rm Ne~II}]12.81\micron~ \approx0.4  \;\;\;\;\;\;\;\;\; 100\%~{\rm H~II} \nonumber \\
{\rm EW}[6.2\micron {\rm PAH}]\approx0.01 \micron,  &&\;\;\; [{\rm Si~II}]34.82\micron / [{\rm S~III}]33.48\micron~~\approx3.5  \;\;\;\;\;\;\;\;\; 100\%~{\rm AGN}  \nonumber \\
{\rm EW}[6.2\micron {\rm PAH}]\approx0.7  \micron,~ &&\;\;\; [{\rm Si~II}]34.82\micron / [{\rm S~III}]33.48\micron~~\approx0.2  \;\;\;\;\;\;\;\;\; 100\%~{\rm H~II} \nonumber
\end{eqnarray}
The dashed line in the upper panel of Figure~\ref{fig:lines_pah} shows the approximate mixing model of Genzel et al. (1998; their Figure~5), obtained after empirically deriving a relation between their 7.7\m\ PAH `strength' (line-to-continuum ratio) and the 6.2\m\ PAH equivalent width.  Many of the high ionization line data presented by Genzel et al. (1998) were upper limits, so it is unsurprising that their original curve lies above our curve (though the discrepancy may also lie in small number statistics).

\subsubsection {Line Ratios of Different Ionization States of the Same Element}

\begin{figure}
 \plotone{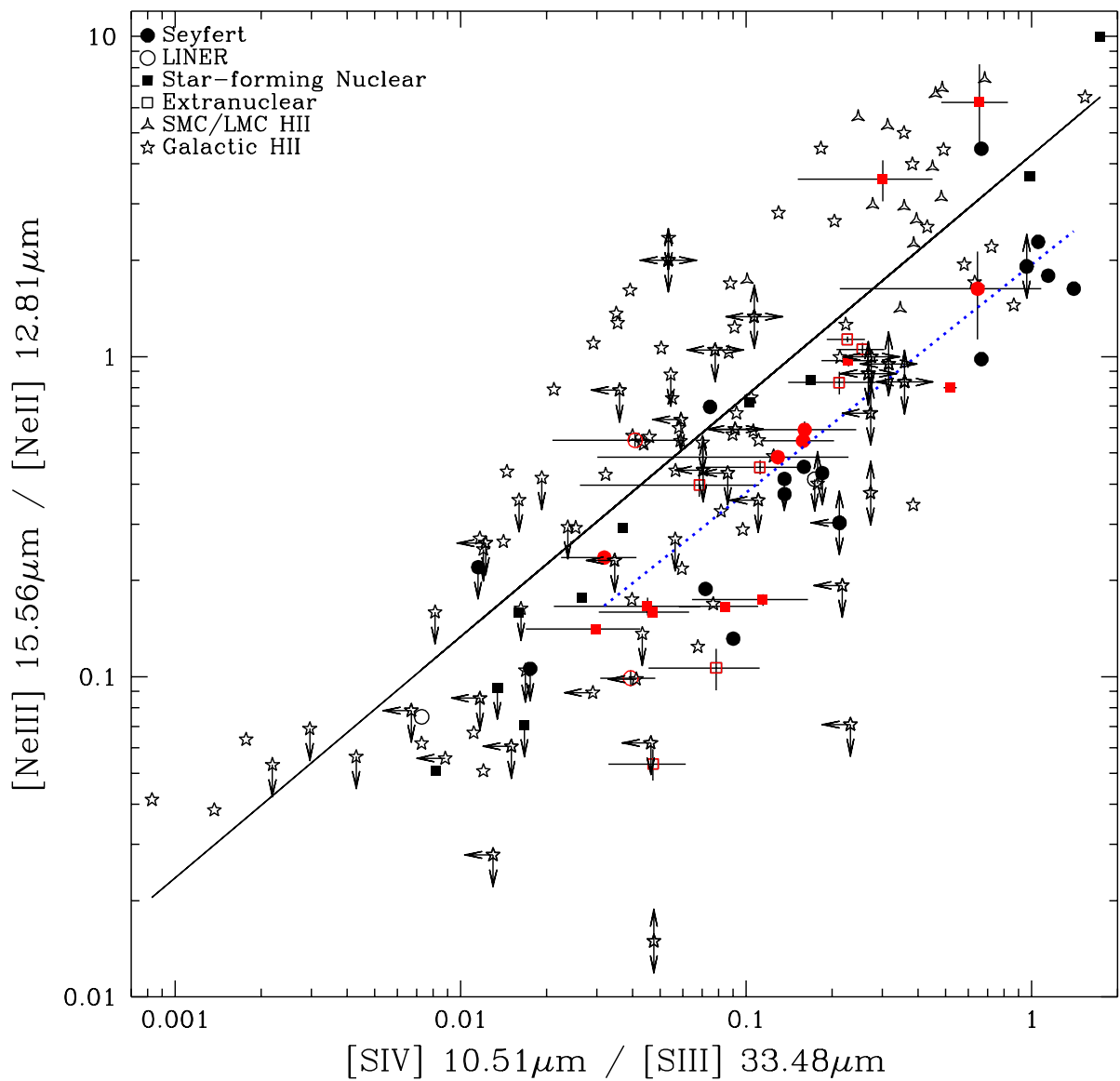}
 \caption{A diagnostic diagram involving ratios of neon and sulfur lines at different ionization levels is displayed (see, for example, Verma et al. 2003).  The data are displayed as described in Figure~\ref{fig:lines_pah}.  The solid line is a linear fit to the detections of star-forming nuclei and \HII\ regions, while the dotted line is linear fit to the Seyfert detections.  }
 \label{fig:NeS}
\end{figure}

Figure~\ref{fig:NeS} plots a ratio of doubly- to singly-ionized neon as a function of a ratio of triply- to doubly-ionized sulfur (see also Verma et al. 2003).  Many of the data points in this plot are for Galactic \HII\ regions (Vermeij et al. 2002; Giveon et al. 2002; Peeters et al. 2002; see \S~\ref{sec:ISO}).  Clearly the neon excitation tracks the sulfur excitation.  To first order, there does not appear to be any sequence in the distribution according to source classification (Seyfert, starburst, \HII, etc).  However, the low-metallicity \HII\ regions from the Magellanic Clouds are preferentially in the high excitation, upper righthand corner of the diagram.  Presumably the diminished line blanketing for low-metallicity sources results in a harder radiation field and thus higher excitations (see Genzel \& Cesarsky 2000; Mart\'in-Hern\'andez et al. 2002; Madden et al. 2006).  In addition, AGN sources show somewhat lower \NeIII 15.6\m/\NeII 12.81\m\ ratios than exhibited by star-forming sources, and the locus of the AGN detections lie at slightly higher values of \SIV 10.51\m/\SIII 33.48\m.  The dotted and solid curves show linear fits to the Seyfert and star-forming sources, and respectively have slopes of 0.71[$\pm$0.12] and 0.75[$\pm$0.06].

\subsubsection {A Neon, Sulfur, and Silicon Diagnostic}
\label{sec:NeSSi}

\begin{figure}
 \plotone{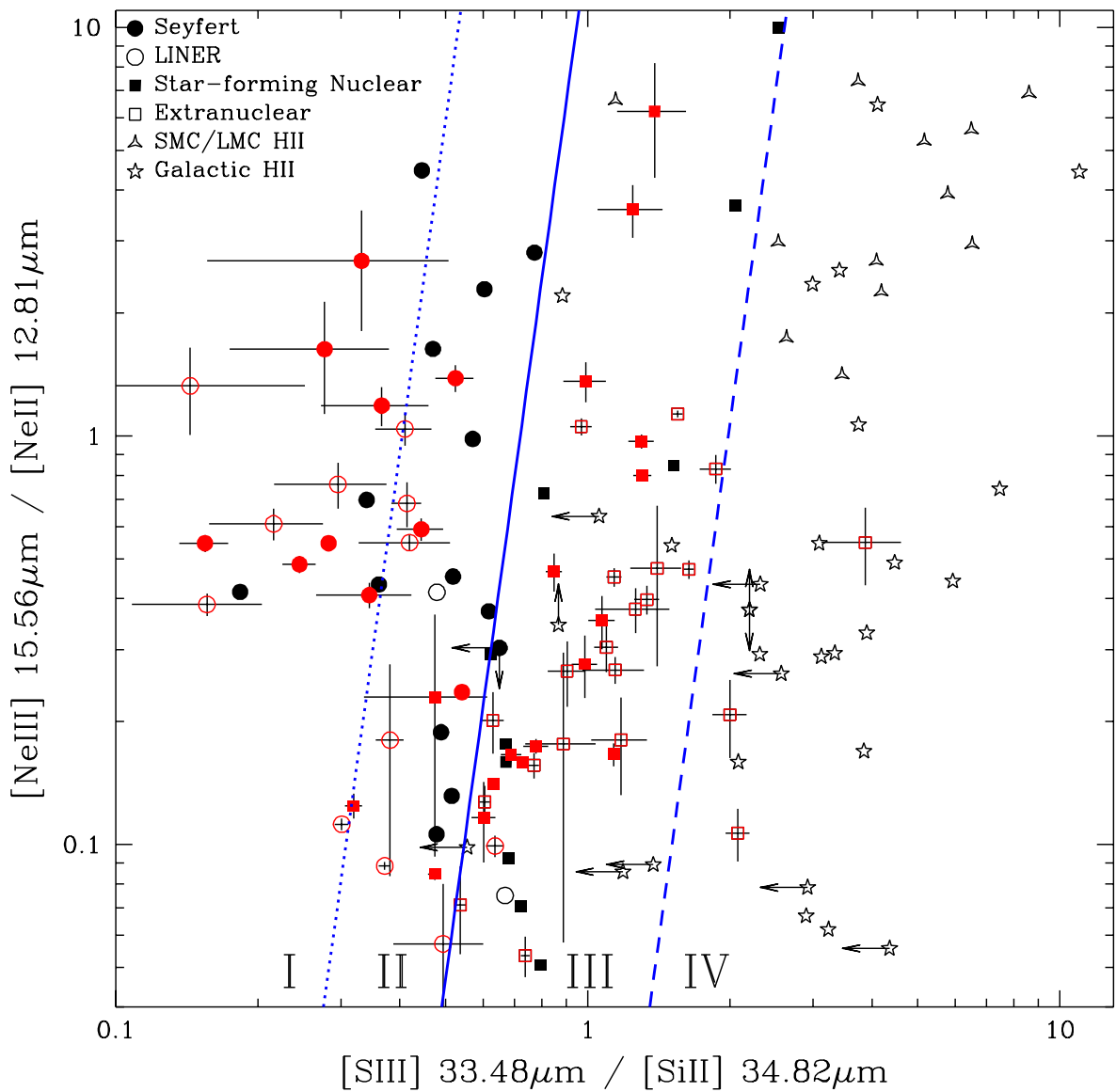}
 \caption{A neon, sulfur, and silicon diagnostic diagram involving ratios of lines at different ionizations is displayed.  The data are displayed as described in Figure~\ref{fig:lines_pah}.  The lines and Roman numerals delineate regions distinguished by Seyferts, LINERs, star formation, etc. (see Table~\ref{tab:NeSSi} and \S~\ref{sec:NeSSi}).}
 \label{fig:NeSSi}
\end{figure}

If the neon excitation is plotted as a function of \SIII 33.48\m/\SiII 34.82\m\ (Figure~\ref{fig:NeSSi}), a more obvious separation of the star-forming and AGN-powered data points is observed.  Not only do the low-metallicity Magellanic Cloud regions exhibit a higher neon excitation, nearly all of the ``pure star-forming'' nuclei and extra-nuclear regions show relatively elevated ratios in \SIII 33.48\m/\SiII 34.82\m\ (see also Figure~\ref{fig:lines_pah}).  Note in addition that many of the filled squares representing starbursting/star-forming nuclei are located between the \HII\ regions and the AGN.  Table~\ref{tab:NeSSi} quantifies the source type fractions within each of the four regions delineated by the lines drawn in Figure~\ref{fig:NeSSi}.  The boundaries are defined by curves with the same slope but differing offsets:
\be
\log ( [{\rm Ne~III}]15.56\micron / [{\rm Ne~II}]12.81\micron ) = 8.4 \log ( [{\rm S~III}]33.48\micron / [{\rm Si~II}]34.82\micron ) + \gamma,
\ee
where $\gamma=$[$+$3.3,$+$1.2,$-$2.5] for the lines demarcating Regions~[I$-$II,II$-$III,III$-$IV].
The numbers provided in Table~\ref{tab:NeSSi} can be used to determine the statistical reliability of a classification for a galaxy randomly drawn from a mid-infrared line survey.  For example, if a galaxy appears in Region~III or Region~IV, it should be classified as a star-forming system with a 1$\sigma$ confidence interval of 84-93\% and 91-98\%, respectively.  Likewise, a galaxy residing in Region~I or Region~II should be classified as AGN-powered with a 1$\sigma$ confidence interval of 83-97\% and 73-88\%, respectively.

These results can be partially understood in the context of the cooling line physics introduced above.  The \SiII 34.82\m\ line is a significant coolant of X-ray ionized regions or dense photodissociation regions (Hollenbach \& Tielens 1999), whereas the \SIII 33.48\m\ line is a strong marker of \HII\ regions.  In other words, extra-nuclear regions and star-forming nuclei will show strong signatures of the Str\"omgren sphere coolant \SIII 33.48\m, while AGN and their associated X-ray dominated regions or dense photodissociation regions will exhibit relatively strong \SiII 34.82\m\ emission in analogy to the increased strength of \OI~6300\AA\ emission in AGN (e.g., Veilleux \& Osterbrock 1987).  In addition, the fraction of photodissociation regions falling within each beam will play a role in the line ratios.  The data for Magellanic Cloud and Galactic \HII\ regions stem from smaller physical apertures and thus are likely to have fractionally higher contributions from Str\"omgren spheres than photodissociation regions.

Metallicity may be a factor as well.  Since the central regions of galaxies typically are more abundant in heavy metals (Pagel \& Edmunds 1981; McCall 1982; Vila-Costas \& Edmunds 1992; Pilyugin \& Ferrini 1998; Henry \& Worthey 1999), and as explained above a lower metallicity can lead to harder radiation fields and thus enhanced high-ionization-to-low-ionization line ratios, it is possible that this AGN$\rightarrow$\HII\ nucleus$\rightarrow$\HII\ region sequencing along the \SIII 33.48\m/\SiII 34.82\m\ axis is affected by metallicity.  However, the lower metallicity Magellanic Cloud data are not substantially to the right of the Galactic \HII\ region data, so the effect is not solely due to metallicity.  
Alternatively, 
perhaps some of the star-forming nuclei have contributions from undetected weak AGN and thus are not ``pure'' star-forming nuclei, resulting in a location for star-forming nuclei on this diagram between AGN and \HII\ regions.

\subsubsection {Density Diagnostics}

\begin{figure}
 \plotone{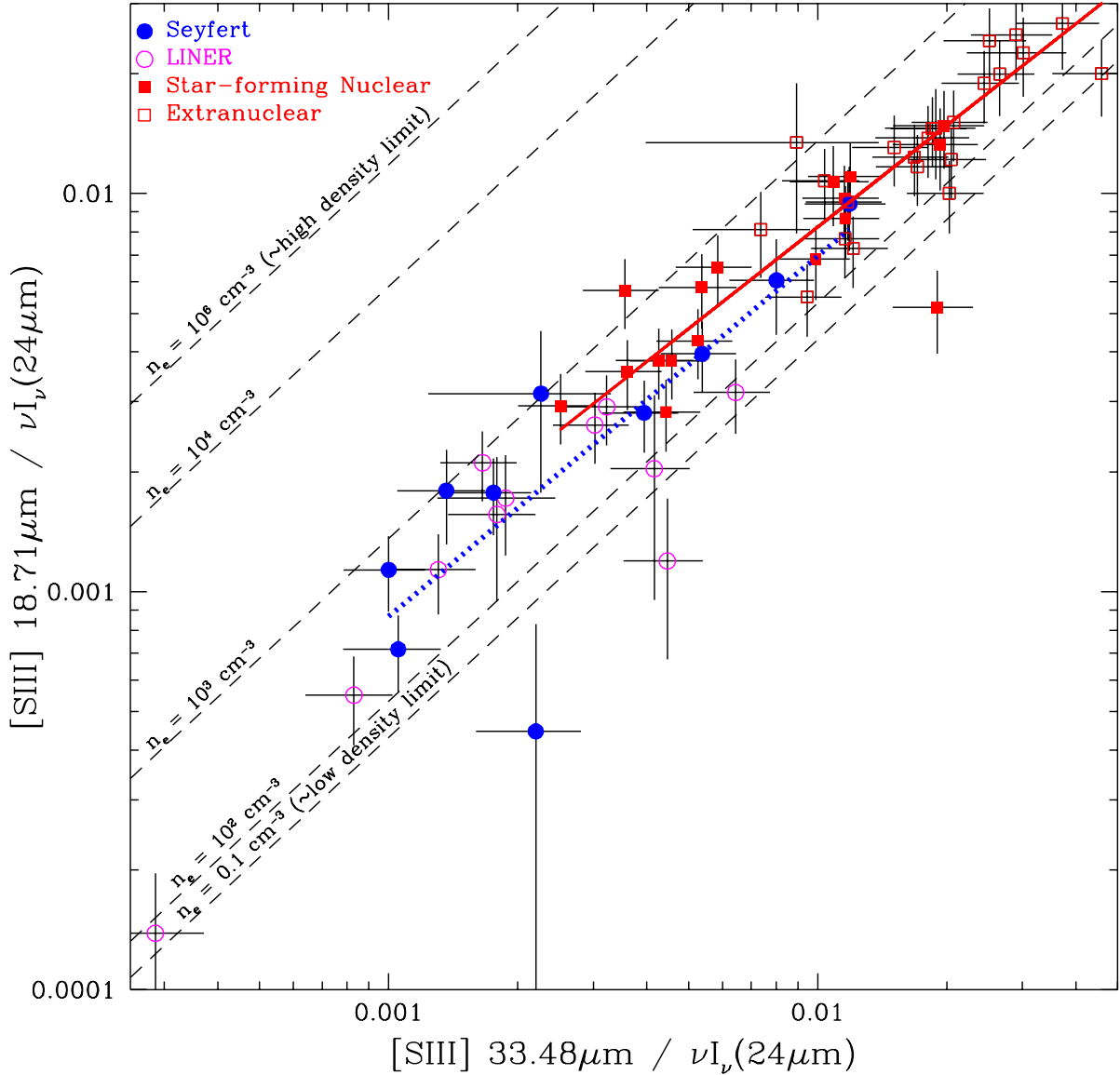}
 \caption{The correlation between two transitions of doubly-ionized sulfur is displayed, normalized by the flux at 24\m.  This plot includes only SINGS data and 1$\sigma$ error bars.  The solid line is a linear fit to the detections of star-forming nuclei and \HII\ regions, while the dotted line is a linear fit to the Seyfert detections.  The set of dashed lines represent different constant interstellar electron densities.  Most of the SINGS data are bounded by the low and high density limiting values, and several are consistent with the low density limiting value of \SIII 18.71\m/\SIII 33.48\m=0.43 at 0.1~cm$^{-3}$. The SINGS data typically exhibit $n_{\rm e}\sim400$~cm$^{-3}$.}
 \label{fig:S}
\end{figure}

The average line ratio of \SIII 18.71\m-to-\SIII 33.48\m\ for the SINGS sample is 0.82 with a 1$\sigma$ dispersion of 0.27.  This ratio implies an interstellar electron density of $\left<n_{\rm e}\right> \sim 400^{+240}_{-290}~{\rm cm}^{-3}$ for the $\sim23$\arcsec$\times$15\arcsec\ nuclear and extranuclear regions of SINGS galaxies.  The average density is calculated using electron collision strengths from Tayal \& Gupta (1999) and excluding the effects of differential extinction at these mid-infrared wavelengths (which is shown in Section~\ref{sec:optical} to be relatively small at optical wavelengths).  This density on $\sim$kiloparsec scales is typical of starburst/LINER/Seyfert galaxies (Kewley et al. 2001), but lower than the $\sim10^3-10^4~{\rm cm}^{-3}$ found for high surface brightness \HII\ regions using small apertures uncontaminated by the surrounding neutral interstellar medium and lower density \HII\ regions (e.g., Wang et al. 2004).  A visual way to explore this doubly-ionized sulfur line ratio is portrayed in Figure~\ref{fig:S} using the aperture-matched 24\m\ flux as a normalization for the line fluxes.  The set of dashed lines represent different interstellar electron densities.  The correlation in Figure~\ref{fig:S} extends over two orders of magnitude in the line-to-continuum ratios and encompasses both AGN-dominated and star-formation-dominated sources; a non-parametric ranking analysis indicates a global correlation at the $7\sigma$ level.  Linear fits to the two separate star-forming and Seyfert populations emphasize that, though the trends for the two populations are similar, the nuclei with Seyfert characteristics (dotted line; slope 0.91$\pm$0.22) differ {\it along the diagonal} from the starbursting nuclei and the \HII\ regions (solid line; slope 0.85$\pm$0.05).  Star-forming sources show more pronounced \SIII 33.48\m-to-continuum ratios compared to Seyferts and LINERs, consistent with the notion that \SIII 33.48\m\ is an important coolant of \HII\ regions.  In addition, we are seeing the effects of continuum dilution in the line-to-continuum ratio for Seyferts and LINERs, sources for which hot dust emission is pronounced in the mid-infrared and thus the line-to-continuum ratios are suppressed (e.g., Laurent et al. 2000; Dale et al. 2001; Xu et al. 2001; Siebenmorgen et al. 2004).  Finally, similar to what is seen in Figures~\ref{fig:lines_pah} and \ref{fig:NeSSi}, but perhaps not as prominently, the data in Figure~\ref{fig:S} suggest that the \HII\ regions (open squares) occupy a different portion of the diagram than star-forming nuclei (filled squares).  \HII\ regions have higher line-to-continuum ratios than star-forming nuclei, which in turn have higher ratios than Seyferts and LINERs.

\section {Summary}

We have presented mid-infrared diagnostic diagrams for a large portion of the SINGS sample supplemented by archival \ISO\ and \Spitzer\ data.  A portion of our work solidifies and extends previous \ISO-based mid-infrared work to lower luminosity normal galaxy nuclei and \HII\ regions using the {\it Spitzer Space Telescope}.  We also present new diagnostics that effectively constrain a galaxy's dominant power source.  The power of the diagnostic diagrams of Genzel et al. (1998; see also Peeters, Spoon, \& Tielens 2004) for distinguishing between AGN and star-forming sources in dusty ULIRGs is that mid-infrared lines and PAH features are much less affected by extinction than their optical counterparts in a traditional diagnostic diagram.
Unlike diagrams 
put forth by Genzel et al., which rely on detecting relatively weak high-ionization lines like \OIV~25.89\m\ and \NeV~14.32\m, we provide a new diagnostic that utilizes a strong {\it low}-ionization line.  The advantage of using a line ratio like \SiII/\NeII\ is that singly-ionized silicon and neon respectively have ionization potentials of only 8.15 and 12.8~eV, so they can both be observed over a large range of physical conditions.
This is similar in concept to previous efforts that have taken advantage of \OI\ lines (e.g., at 6300~\AA\ or 63\m) that are coolants of the X-ray dominated regions (or dense photodissociation regions) surrounding AGN.  In plots of \OIV/\NeII, \SiII/\NeII, and \SiII/\SIII\ vs. 6.2\m\ PAH equivalent width, we identify regions where $>$90\% of the sources are Seyfert or LINER.  Likewise, additional regions in all three plots show populations comprised of more than 90\% \HII\ regions or star-forming nuclei.

Another useful mid-infrared diagnostic is \NeIII~15.56\m/\NeII~12.81\m\ vs. \SIII~33.48\m/\SiII~34.82\m.  This plot tracks the excitation power of the radiation field on one axis, while the other axis is a relative measure of the cooling of \HII\ regions and X-ray dominated regions (or dense photodissociation regions).  Similar to what is found for the diagnostics mentioned above, both starbursting nuclei and extranuclear regions stand apart from nuclei that are powered by accretion-powered disks.  Moreover, compared to starbursting nuclei, extranuclear regions typically separate even further from Seyfert nuclei, especially for low-metallicity environments.  Presumably this extranuclear$\longleftrightarrow$nuclear separation occurs since extranuclear regions are cleaner representatives of \HII\ regions than starburst nuclei, because their stellar populations and interstellar medium structure are less complex.  Extranuclear regions more likely contain younger stellar populations since they trace a single burst, as opposed to the average of multiple star formation episodes for nuclei (e.g., Dale et al. 2004b).  Finally, we note that it is difficult to clearly distinguish between pure Seyfert and pure LINER sources using these diagnostics. 

The line ratio \SIII~18.71\m/\SIII~33.48\m\ yields an average interstellar electron density of $\left<n_{\rm e}\right> \sim 400^{+240}_{-290}~{\rm cm}^{-3}$ for the $\sim23$\arcsec$\times$15\arcsec\ nuclear and extranuclear regions of SINGS galaxies.  This density is much closer (in log space) to the theoretical low density limit of Tayal \& Gupta (1999) than their high density limit, and in fact the data for several sources are consistent with the low density limiting value.  In addition to the interstellar gas densities being unremarkable, there are no SINGS sources sufficiently obscured by dust such that optical and infrared diagnostics provide obviously discrepant classifications of the energy source.  This is not surprising, however, since our nuclei exhibit modest extinctions, $\left< A_V \right> \sim 1.0$~mag, and normal star-forming galaxy nuclei in general show $0\lesssim A_V \lesssim 3$~mag (Ho, Filippenko, \& Sargent 1997b).  This relative transparency means the SINGS sample is not ideally suited for a detailed comparison of the relative merits of optical and infrared classifications.  However, the diverse SINGS sample of nuclear and extranuclear regions provides an enormous range of physical parameters, and this has proved critical in developing the mid-infrared diagnostics in this paper.

\acknowledgements 
Mike Brotherton provided helpful comments.  Support for this work, part of the {\it Spitzer Space Telescope} Legacy Science Program, was provided by NASA through Contract Number 1224769 issued by the Jet Propulsion Laboratory, California Institute of Technology under NASA contract 1407.  This research has made use of the NASA/IPAC Extragalactic Database which is operated by JPL/Caltech, under contract with NASA.  This publication makes use of data products from the Two Micron All Sky Survey, which is a joint project of the University of Massachusetts and the Infrared Processing and Analysis Center/California Institute of Technology, funded by the National Aeronautics and Space Administration and the National Science Foundation.


\begin{deluxetable}{lcccccccc}
\def\a{\tablenotemark{a}}

\def\p{$\pm$}
\tabletypesize{\scriptsize}
\tablenum{1}
\label{tab:fluxes_nuc}
\tablecaption{Nuclear Emission Line Fluxes and 6.2\m\ PAH Feature Equivalent Widths}
\tablewidth{0pc}
\tablehead{
\colhead{Species} &
\colhead{PAH} &
\colhead{[SIV]} &
\colhead{[NeII]} &
\colhead{[NeIII]} &
\colhead{[SIII]} &
\colhead{[OIV]} &
\colhead{[SIII]} &
\colhead{[SiII]} 
\\
\colhead{Wavelength} &
\colhead{6.2$\mu$m} &
\colhead{10.51$\mu$m} &
\colhead{12.81$\mu$m} &
\colhead{15.56$\mu$m} &
\colhead{18.71$\mu$m} &
\colhead{25.89$\mu$m} &
\colhead{33.48$\mu$m} &
\colhead{34.82$\mu$m} 
\\
\colhead{ionization (eV)} & 
\colhead{} &
\colhead{34.8} &
\colhead{21.6} &
\colhead{41.0} &
\colhead{23.3} &
\colhead{54.9} &
\colhead{23.3} &
\colhead{8.2} 
}
\startdata
NGC~0337         &0.56\p0.04&\nodata  &17.4\p1.6&~8.1\p0.5&13.3\p0.6&~0.5\p0.2&17.9\p0.5&21.1\p0.6\cr
NGC~0584         &\nodata   &\nodata  &\nodata  &~2.4\p0.7&~1.1\p0.5&~$<$0.8~~&\nodata  &\nodata  \cr
NGC~0628         &0.39\p0.08&\nodata  &~6.4\p1.3&\nodata  &~2.5\p0.3&~$<$0.7~~&~9.1\p0.7&~5.8\p0.8\cr
NGC~1097         &0.42\p0.01&\nodata  &328\p4~~~&29.0\p0.6&89.6\p0.5&~$<$5.3~~&103\p3~~~&278\p4~~~\cr
NGC~1266         &0.29\p0.02&\nodata  &28.0\p1.0&10.8\p0.6&~1.3\p0.5&~$<$2.8~~&~2.7\p0.6&17.3\p3.7\cr
NGC~1377         &\nodata   &\nodata  &~3.5\p0.5&\nodata  &\nodata  &~$<$3.4~~&\nodata  &\nodata  \cr
NGC~1404         &\nodata   &\nodata  &~1.9\p0.8&~1.3\p0.6&\nodata  &~$<$0.7~~&\nodata  &\nodata  \cr
NGC~1566         &0.20\p0.01&~1.1\p0.6&16.6\p0.8&~9.8\p0.4&~6.8\p0.6&~6.7\p0.5&~6.8\p0.7&15.2\p0.8\cr
NGC~1705         &0.42\p0.26&~3.2\p0.8&~1.1\p0.3&~7.0\p0.4&~3.4\p0.4&~1.2\p0.4&~4.9\p0.5&~3.6\p0.5\cr
NGC~2798         &0.52\p0.02&~5.9\p1.8&208\p2~~~&34.4\p0.5&81.6\p0.8&~9.6\p1.8&70.1\p2.6&102\p4~~~\cr
NGC~2841         &0.03\p0.03&\nodata  &~5.5\p0.5&~6.6\p0.4&~2.7\p0.5&~0.9\p0.1&~3.6\p0.4&~9.7\p2.3\cr
NGC~2915         &\nodata   &~1.8\p0.9&~3.1\p0.4&11.1\p0.4&~4.6\p0.4&~0.4\p0.1&~6.1\p0.8&~4.9\p0.4\cr
NGC~2976         &0.35\p0.04&\nodata  &~7.7\p0.7&~2.7\p0.3&~6.3\p0.4&~0.4\p0.1&~9.2\p0.3&~8.5\p0.5\cr
NGC~3049         &0.62\p0.04&~1.1\p0.6&36.4\p0.8&~6.1\p0.4&27.3\p0.6&~$<$0.7~~&24.4\p0.4&21.5\p0.5\cr
NGC~3190         &0.09\p0.01&\nodata  &~7.4\p0.6&~5.6\p0.5&~1.8\p0.6&~0.7\p0.2&~2.0\p0.2&~6.9\p1.7\cr
NGC~3184         &0.26\p0.04&\nodata  &17.8\p0.5&~2.1\p0.5&~7.8\p0.5&~$<$0.7~~&~7.2\p0.3&12.0\p0.4\cr
NGC~3198         &0.31\p0.03&\nodata  &13.9\p0.4&~0.8\p0.3&~4.8\p0.5&~$<$0.9~~&~5.5\p0.5&11.1\p2.2\cr
NGC~3265         &0.57\p0.03&~1.5\p0.7&28.4\p0.3&~4.9\p0.2&11.7\p0.3&~$<$1.2~~&13.1\p0.6&16.9\p0.7\cr
Markarian~33     &0.51\p0.02&17.1\p0.8&56.7\p1.1&45.3\p0.4&52.8\p0.7&~1.0\p0.7&32.8\p0.8&25.1\p0.9\cr
NGC~3351         &0.46\p0.01&\nodata  &165\p2~~~&14.0\p0.4&60.0\p0.5&~$<$2.3~~&60.5\p1.7&128\p2~~~\cr
NGC~3521         &0.38\p0.04&\nodata  &12.9\p0.8&~7.9\p0.5&~3.3\p0.7&~0.8\p0.4&~3.6\p0.8&16.7\p2.3\cr
NGC~3627         &0.25\p0.01&\nodata  &21.9\p0.9&~8.9\p0.5&~4.7\p0.4&~1.3\p0.7&~7.0\p1.1&20.2\p3.3\cr
NGC~3773         &0.65\p0.08&~3.9\p0.7&16.5\p0.5&16.1\p0.5&14.3\p0.6&~$<$0.8~~&17.0\p0.6&13.1\p0.7\cr
NGC~3938         &0.13\p0.03&\nodata  &~5.7\p0.5&~1.0\p0.5&~1.1\p0.5&~0.3\p0.1&~2.2\p0.1&~5.9\p0.2\cr
NGC~4125         &\nodata   &\nodata  &~2.5\p0.5&~3.4\p0.5&\nodata  &~$<$0.9~~&~0.8\p0.6&~5.4\p1.1\cr
NGC~4254         &0.54\p0.03&\nodata  &51.7\p1.1&~6.4\p0.4&10.0\p0.5&~1.3\p0.2&15.6\p0.6&48.9\p1.1\cr
NGC~4321         &0.49\p0.02&\nodata  &110\p2~~~&12.3\p0.4&23.1\p0.3&~$<$1.4~~&25.5\p0.8&84.8\p0.9\cr
NGC~4536         &0.54\p0.01&~7.5\p2.6&307\p3~~~&48.9\p0.7&130\p1~~~&~$<$3.9~~&160\p3~~~&220\p3~~~\cr
NGC~4552         &\nodata   &\nodata  &~1.8\p0.5&~4.8\p0.8&~0.9\p0.4&~$<$0.9~~&~0.7\p0.3&~2.0\p0.7\cr
NGC~4569         &0.26\p0.02&~1.0\p0.8&32.6\p1.2&15.8\p0.5&~8.8\p0.7&~2.4\p0.6&~7.7\p0.6&31.5\p0.9\cr
NGC~4579         &0.09\p0.02&\nodata  &22.8\p0.7&12.5\p0.5&~4.0\p0.7&~2.2\p0.3&~3.0\p0.4&19.5\p0.4\cr
NGC~4725         &0.07\p0.04&~0.8\p0.5&~1.8\p0.5&~3.0\p0.4&~0.3\p0.2&~1.7\p0.2&~1.3\p0.2&~4.6\p1.5\cr
NGC~4736         &0.16\p0.01&\nodata  &13.7\p1.1&14.3\p0.5&~6.1\p0.9&~3.8\p0.9&~9.2\p1.0&22.3\p1.8\cr
NGC~4826         &0.28\p0.01&~1.8\p0.5&99.0\p1.6&23.4\p0.6&41.8\p0.4&~2.5\p1.1&56.8\p1.0&105\p1~~~\cr
NGC~5194         &0.24\p0.02&~2.9\p0.8&66.2\p0.8&36.2\p0.6&13.1\p0.6&14.9\p1.6&18.3\p0.4&64.8\p0.8\cr
NGC~5408\a       &\nodata   &\nodata  &\nodata  &~1.7\p0.4&\nodata  &~$<$0.9~~&~1.5\p0.5&~1.8\p0.5\cr
NGC~5713         &0.56\p0.02&~1.7\p0.7&123\p2~~~&17.3\p0.4&47.5\p0.5&~2.0\p1.0&57.2\p1.2&90.3\p1.2\cr
NGC~5866         &0.07\p0.01&\nodata  &~7.5\p0.9&~5.1\p0.3&~1.1\p0.4&~0.7\p0.1&~4.0\p0.3&~9.7\p0.4\cr
IC~4710          &\nodata   &~3.6\p0.8&\nodata  &~4.3\p1.6&~3.6\p0.5&~$<$1.3~~&~4.2\p0.6&~1.4\p0.4\cr
NGC~7331         &0.09\p0.02&~0.5\p0.2&18.8\p0.6&10.3\p0.3&~6.0\p0.4&~2.4\p0.6&12.2\p0.3&29.0\p6.3\cr
NGC~7552         &0.45\p0.01&~6.5\p1.4&573\p31~~&56.8\p1.4&211\p1~~~&~$<$15.3~&165\p5~~~&260\p8~~~\cr
NGC~7793         &0.36\p0.04&\nodata  &10.3\p0.6&~2.9\p0.5&~8.7\p0.6&~$<$0.5~~&~9.4\p0.3&~9.5\p0.5
\enddata
\tablecomments{\footnotesize Fluxes and their (statistical) uncertainties are averaged over $\sim$23\arcsec$\times$15\arcsec\ and listed in units of 10$^{-9}$~W~m$^{-2}$~sr$^{-1}$.  Calibration uncertainties are an additional $\sim$30\%.  The 6.2\m\ PAH feature equivalent width is given in units of microns.}
\tablecomments{\footnotesize Eight galaxies from the SINGS Data Release Three are not listed in this table.  No Short-Low, Short-High, or Long-High data were taken for the optical centers of Holmberg~II, IC~2574, DDO~154 and NGC~6822.  DDO~053, Holmberg~IX, DDO~165, and NGC~5398 (Tololo~89)\a were non-detections.}
\tablenotetext{a}{\footnotesize The infrared emission peaks outside of the field of view of the spectral maps.}
\end{deluxetable}

\begin{deluxetable}{lrrrrrrrr}
\def\a{\tablenotemark{a}}

\def\p{$\pm$}
\tabletypesize{\scriptsize}
\tablenum{2}
\label{tab:fluxes_extranuc}
\tablecaption{Extranuclear Emission Line Fluxes and 6.2\m\ PAH Feature Equivalent Widths}
\tablewidth{0pc}
\tablehead{
\colhead{Species} &
\colhead{PAH} &
\colhead{[SIV]} &
\colhead{[NeII]} &
\colhead{[NeIII]} &
\colhead{[SIII]} &
\colhead{[OIV]} &
\colhead{[SIII]} &
\colhead{[SiII]} 
\\
\colhead{Wavelength} &
\colhead{6.2$\mu$m} &
\colhead{10.51$\mu$m} &
\colhead{12.81$\mu$m} &
\colhead{15.56$\mu$m} &
\colhead{18.71$\mu$m} &
\colhead{25.89$\mu$m} &
\colhead{33.48$\mu$m} &
\colhead{34.82$\mu$m} 
\\
\colhead{ionization (eV)} & 
\colhead{} &
\colhead{34.8} &
\colhead{21.6} &
\colhead{41.0} &
\colhead{23.3} &
\colhead{54.9} &
\colhead{23.3} &
\colhead{8.2} 
}
\startdata
NGC~5194~CCM107 &0.47\p0.02&\nodata  &32.0\p0.6&~2.3\p0.6&10.6\p0.5&~1.7\p0.3&18.3\p0.5&34.1\p0.5\cr
NGC~5194~CCM072 &0.61\p0.02&~1.7\p0.5&71.9\p1.0&~3.8\p0.4&29.4\p0.6&~1.4\p0.8&36.2\p0.5&49.2\p0.7\cr
NGC~5194~CCM071 &0.66\p0.02&\nodata  &44.3\p0.7&~5.6\p0.5&16.6\p0.7&~1.3\p0.4&25.0\p0.6&41.2\p0.5\cr
NGC~5194~CCM001 &0.69\p0.02&\nodata  &16.4\p1.1&~3.3\p0.5&~8.2\p0.3&~0.6\p0.2&13.7\p0.5&21.7\p0.9\cr
NGC~5194~CCM010 &0.69\p0.03&\nodata  &39.2\p0.7&~6.1\p0.4&19.8\p0.5&~0.5\p0.3&29.0\p0.5&37.7\p0.7\cr
NGC~5194~CCM071A&0.52\p0.02&\nodata  &21.3\p0.6&~5.7\p0.4&11.0\p0.5&~$<$0.6~~&14.4\p2.1&12.6\p0.4\cr
NGC~3031~HK230  &0.44\p0.03&\nodata  &~4.1\p0.5&~0.7\p0.5&~2.7\p0.4&~$<$0.7~~&~3.6\p0.5&~4.0\p0.4\cr
NGC~3031~HK343  &0.38\p0.03&\nodata  &~7.5\p3.1&~3.6\p0.4&~7.0\p0.9&~$<$0.8~~&~9.8\p0.8&~7.0\p0.6\cr
NGC~3031~HK453  &0.65\p0.05&\nodata  &~9.0\p0.6&~3.4\p0.4&~8.8\p0.9&~$<$0.8~~&11.7\p2.0&~9.3\p0.6\cr
NGC~3031~HK268  &0.55\p0.03&~1.2\p0.7&15.1\p0.8&~6.0\p0.4&13.4\p0.7&~$<$0.8~~&17.3\p0.7&13.0\p0.6\cr
NGC~3031~HK652  &0.68\p0.06&\nodata  &10.3\p0.7&~2.7\p0.5&~9.2\p1.1&~$<$0.8~~&10.7\p0.8&11.8\p0.6\cr
NGC~3031~HK741  &0.75\p0.05&\nodata  &~8.9\p0.8&~1.6\p0.4&~7.3\p0.4&~$<$0.9~~&~7.6\p0.7&~6.5\p0.7\cr
NGC~3031~Munch1 &2.98\p2.29&\nodata  &\nodata  &~1.4\p0.2&\nodata  &~$<$1.0~~&\nodata  &\nodata  \cr
NGC~6946~H4     &0.65\p0.02&~3.6\p0.7&16.6\p0.6&17.5\p0.5&14.7\p0.4&~1.1\p0.2&14.2\p0.6&14.6\p0.5\cr
NGC~6946~HK3    &0.63\p0.03&~8.8\p1.3&28.2\p0.4&31.9\p0.4&30.6\p0.4&~0.9\p0.2&39.0\p0.5&25.1\p0.5\cr
NGC~6946~H288   &0.81\p0.04&\nodata  &18.1\p0.4&~8.5\p0.4&14.0\p0.5&~$<$0.6~~&19.3\p0.3&11.8\p0.2\cr
NGC~6946~H40    &0.84\p0.03&~1.9\p0.8&18.8\p0.4&~8.5\p0.4&15.0\p0.4&~0.2\p0.1&17.3\p0.3&15.2\p0.5\cr
NGC~6946~H28    &1.05\p0.45&\nodata  &~9.4\p0.5&~2.9\p0.3&~6.8\p0.5&~$<$0.5~~&~9.0\p0.3&~8.2\p0.4\cr
NGC~0628~H292   &0.55\p0.02&~2.1\p0.9&25.6\p0.7&~2.7\p0.4&20.1\p0.6&~$<$1.0~~&27.4\p0.5&13.1\p0.7\cr
NGC~0628~H572   &0.69\p0.03&\nodata  &12.5\p0.7&~2.6\p0.5&~8.1\p0.4&~$<$0.9~~&16.3\p0.5&~8.2\p0.6\cr
NGC~0628~H627   &0.60\p0.03&~3.8\p1.3&11.5\p0.7&~9.5\p0.5&10.6\p0.4&~$<$0.9~~&17.8\p0.9&~9.6\p0.5\cr
NGC~0628~H013   &0.52\p0.07&\nodata  &~4.1\p0.7&~2.2\p0.3&~4.9\p0.4&~$<$0.7~~&11.3\p1.3&~2.9\p0.4\cr
HolmbergII~HSK45&\nodata   &\nodata  &\nodata  &~3.4\p0.5&~3.2\p0.4&~$<$0.9~~&~2.9\p0.7&~2.8\p0.5\cr
HolmbergII~HSK67&\nodata   &\nodata  &\nodata  &~0.7\p0.2&\nodata  &~$<$0.7~~&~0.6\p0.2&~0.9\p0.3\cr
HolmbergII~HSK70&\nodata   &\nodata  &\nodata  &~1.1\p0.5&~0.6\p0.2&~$<$0.9~~&~0.4\p0.2&~1.7\p0.5\cr
HolmbergII~HSK07&\nodata   &\nodata  &\nodata  &~1.1\p0.6&~0.8\p0.5&~$<$1.0~~&\nodata  &\nodata  
\enddata
\tablecomments{\footnotesize Fluxes and their (statistical) uncertainties  are averaged over $\sim$23\arcsec$\times$15\arcsec\ and listed in units of 10$^{-9}$~W~m$^{-2}$~sr$^{-1}$.  Calibration uncertainties are an additional $\sim$30\%.  The 6.2\m\ PAH feature equivalent width is given in units of microns.}
\end{deluxetable}

\begin{deluxetable}{lll}
\def\a{\tablenotemark{a}}

\def\p{$\pm$}
\tabletypesize{\scriptsize}
\tablenum{3}
\label{tab:archival}
\tablecaption{Archival Sources}
\tablewidth{0pc}
\tablehead{
\colhead{Object} &
\colhead{Type} &
\colhead{Reference} 
}
\startdata
LMC~N160A1       &LMC HII      &Vermeij et al. (2002)    \cr
LMC~N160A2       &LMC HII      &Vermeij et al. (2002)    \cr
LMC~N159-5       &LMC HII      &Vermeij et al. (2002)    \cr
LMC~N157B        &LMC HII      &Vermeij et al. (2002)    \cr
LMC~N4A          &LMC HII      &Vermeij et al. (2002)    \cr
LMC~N11A         &LMC HII      &Vermeij et al. (2002)    \cr
LMC~N83B         &LMC HII      &Vermeij et al. (2002)    \cr
LMC~30Dor1       &LMC HII      &Vermeij et al. (2002)    \cr
LMC~30Dor2       &LMC HII      &Vermeij et al. (2002)    \cr
LMC~30Dor3       &LMC HII      &Vermeij et al. (2002)    \cr
LMC~30Dor4       &LMC HII      &Vermeij et al. (2002)    \cr
SMC~N88A         &SMC HII      &Vermeij et al. (2002)    \cr
SMC~N66          &SMC HII      &Vermeij et al. (2002)    \cr
SMC~N81          &SMC HII      &Vermeij et al. (2002)    \cr
NGC~0253         &HII nucleus  &Verma et al. (2003)    \cr
IC~342           &HII nucleus  &Verma et al. (2003)    \cr
II~Zw~40         &HII nucleus  &Verma et al. (2003)    \cr
NGC~3034         &HII nucleus  &Verma et al. (2003)    \cr
NGC~3256         &HII nucleus  &Verma et al. (2003)    \cr
NGC~3690A        &HII nucleus  &Verma et al. (2003)    \cr
NGC~3690B        &HII nucleus  &Verma et al. (2003)    \cr
NGC~4038         &HII nucleus  &Verma et al. (2003)    \cr
NGC~4945         &HII/Seyfert  &Verma et al. (2003)    \cr
NGC~5236         &HII nucleus  &Verma et al. (2003)    \cr
NGC~5253         &HII nucleus  &Verma et al. (2003)    \cr
NGC~7552         &LINER/HII    &Verma et al. (2003)    \cr
WB89~380~A       &Galactic HII &Giveon et al. (2002)    \cr
WB89~380~B       &Galactic HII &Giveon et al. (2002)    \cr
WB89~399         &Galactic HII &Giveon et al. (2002)    \cr
\enddata                               
\tablecomments{\footnotesize The full version would appear as an electronic table in the online Journal.}
\end{deluxetable}

\begin{deluxetable}{cccccc}
\def\a{\tablenotemark{a}}

\def\p{$\pm$}
\tabletypesize{\scriptsize}
\tablenum{4}
\label{tab:lines_pah}
\tablecaption{Classifications by Region in Figure~\ref{fig:lines_pah}}
\tablewidth{0pc}
\tablehead{
\colhead{Region} &
\colhead{Number} &
\colhead{Seyfert} &
\colhead{LINER} &
\colhead{HII} &
\colhead{extranuclear} 
\\
\colhead{} &
\colhead{of} &
\colhead{} &
\colhead{} &
\colhead{nuclei} &
\colhead{+HII regions} 
\\
\colhead{} &
\colhead{Sources} &
\colhead{(\%)} &
\colhead{(\%)} &
\colhead{(\%)} &
\colhead{(\%)} 
}
\startdata
I     &15&73&20&~7&~~0\cr
II    &31&42&23&19&~16\cr
III   &31&~0&10&32&~58\cr
\hline
IV    &12&67&25&~8&~~0\cr
V     &39&33&18&21&~28\cr
VI    &34&~0&~9&26&~65\cr
\hline
VII   &12&67&25&~8&~~0\cr
VIII  &45&31&24&27&~18\cr
IX    &31&~0&~0&16&~84\cr
\enddata                               
\tablecomments{\footnotesize ``extranuclear + HII regions'' implies SINGS extranuclear/HII regions in addition to Milky Way and Magellanic Cloud HII regions.}
\end{deluxetable}

\begin{deluxetable}{cccccc}
\def\a{\tablenotemark{a}}

\def\p{$\pm$}
\tabletypesize{\scriptsize}
\tablenum{5}
\label{tab:NeSSi}
\tablecaption{Classifications by Region in Figure~\ref{fig:NeSSi}}
\tablewidth{0pc}
\tablehead{
\colhead{Region} &
\colhead{Number} &
\colhead{Seyfert} &
\colhead{LINER} &
\colhead{HII} &
\colhead{extranuclear} 
\\
\colhead{} &
\colhead{of} &
\colhead{} &
\colhead{} &
\colhead{nuclei} &
\colhead{+HII regions} 
\\
\colhead{} &
\colhead{Detections} &
\colhead{(\%)} &
\colhead{(\%)} &
\colhead{(\%)} &
\colhead{(\%)} 
}
\startdata
I+II  &38&61&32&~8&~~0\cr
III+IV&79&~0&~3&30&~67\cr
\hline
I     &16&69&31&~0&~~0\cr
II    &22&55&32&14&~~0\cr
III   &47&~0&~4&51&~45\cr
IV    &32&~0&~0&~0&100\cr
\enddata                               
\tablecomments{\footnotesize ``extranuclear + HII regions'' implies SINGS extranuclear/HII regions in addition to Milky Way and Magellanic Cloud HII regions.}
\end{deluxetable}

\end{document}